\documentclass[prc,twocolumn,floatfix,groupedaddress,nofootinbib,showpacs,preprintnumbers,
amsmath,amssymb,amsfonts,superscriptaddress,widetable] {revtex4-1}
\usepackage{graphicx}
\usepackage{dcolumn}
\usepackage{mathrsfs}
\usepackage{bm}
\usepackage{float}
\usepackage{soul}
\usepackage{relsize}
\usepackage[usenames]{color}

\newcommand{\alphaD}{$\alpha_{\raisebox{-1pt}{\tiny D}}$}

\begin{document}

\title{On the possible existence of a soft dipole mode in ${}^{8}{\rm He}$}
\author{J. Piekarewicz}
\email{jpiekarewicz@fsu.edu}
\affiliation{Department of Physics, Florida State University, Tallahassee, FL 32306}
\date{\today}
\begin{abstract}
 With an extreme neutron-to-proton ratio of $N/Z\!=\!3$, ${}^{8}{\rm He}$ 
 provides an ideal laboratory for the study of a variety of exotic phenomena, 
 such as the emergence of a soft dipole mode that is dominated by transitions 
 into the continuum. In this contribution, a covariant density functional theory 
 (DFT) framework is used to compute ground-state properties and the dipole 
 response of ${}^{8}{\rm He}$. Although ${}^{8}{\rm He}$ is admittedly too light 
 for DFT to be applicable, the great merit of the approach is that the spurious 
 contamination associated with the center-of-mass motion is guaranteed to 
 decouple from the physical response. Given that a strong mixing between the 
 isoscalar and isovector dipole modes is expected for a system with such a 
 large neutron-proton asymmetry as ${}^{8}{\rm He}$, the narrow structures 
 that emerged at low energies in the isovector dipole response are attributed 
 to the shift of the spurious strength to zero (or near zero) excitation energy. 
 Thus, the theoretical framework implemented here disfavors the emergence 
 of a soft dipole mode in ${}^{8}{\rm He}$.
\end{abstract}

\maketitle

\section{Introduction}
\label{introduction}
What combinations of neutrons and protons can form a bound atomic 
nucleus is one of the overarching questions animating nuclear science 
today\,\cite{LongRangePlan}. A core mission of nuclear science is to map 
the neutron drip line, which requires the identification of the most 
neutron-rich element in an isotopic chain that remains stable against 
particle decay. So far, the neutron drip line has been mapped up to an 
including fluorine and neon\,\cite{Ahn:2019xgh}---a challenging 
experimental task that took almost two decades since the confirmation 
of ${}^{24}{\rm O}$ as the $Z\!=\!8$ dripline 
nucleus\,\cite{Guillemaud-Mueller:1990vpa,Hoffman:2009zza}.
In the case of helium, the last stable isotope is ${}^{8}{\rm He}$---an exotic 
nucleus with an extreme neutron-to-proton ratio of $N/Z\!=\!3$; see 
Ref.\cite{Thoennessen:2004} and references contained therein. 
Among the novel behavior that emerges at the limits of stability is the 
development of neutron halos and neutron skins, due to either a 
low neutron separation energy or a large neutron-proton asymmetry. 
Besides the development of extended spatial distributions, weakly 
bound nuclei often give rise to soft modes of excitation that involve 
transitions into the continuum. 

An early experiment using the Coulomb excitation of ${}^{8}{\rm He}$
identified a soft dipole resonance at an excitation energy of about 
4\,MeV\,\cite{Markenroth:2001oxa,Meister:2002uuu}. Later on, Golovkov, 
Grigorenko, and collaborators populated the low-lying spectrum of 
${}^{8}{\rm He}$ via a transfer reaction and confirmed the existence 
of a soft dipole mode, albeit at a slightly lower energy of about 
3\,MeV\,\cite{Golovkov:2008kg,Grigorenko:2009}. In contrast, one
of the main findings of the dissociation experiment on 
${}^{8}{\rm He}$ performed at Michigan State University concluded 
that an insignificant fraction of no more than 3\% of the energy weighted 
sum rule is exhausted by the low-energy mode\,\cite{Iwata:2000dd}. 
This result has been validated by the recent inelastic proton scattering 
experiment that concluded that the measured angular distribution is not 
consistent with a dipole excitation\,\cite{Holl:2021bxg}. It is anticipated 
that the high statistics experiment already finalized at the RIKEN facility 
in Japan will settle the issue\,\cite{Aumann:2022PC}.

From the theoretical perspective, ground-state properties of 
${}^{8}{\rm He}$ have been computed using a variety of state-of-the-art 
ab initio methods\,\cite{Bacca:2009yk,Caprio:2014iha,Holl:2021bxg,Bonaiti:2021kkp}.
However, to our knowledge, it is only the very recent ab initio work by Bonaiti, 
Bacca, and Hagen\,\cite{Bonaiti:2021kkp} that addresses the possible 
existence of a soft dipole mode in ${}^{8}{\rm He}$. The authors have merged
the coupled-cluster framework to the Lorentz-integral-transform 
approach\,\cite{Bacca:2013dma} to report on the emergence of low-energy dipole 
strength around 5\,MeV, in agreement with 
Refs.\cite{Markenroth:2001oxa,Meister:2002uuu,Golovkov:2008kg,Grigorenko:2009}, 
but in disagreement with Refs.\cite{Iwata:2000dd,Holl:2021bxg}.

In this paper we offer an alternative theoretical perspective based on 
density functional theory. Density Functional Theory (DFT) is a powerful technique 
developed by Kohn and collaborators\,\cite{Hohenberg:1964zz,Kohn:1965}, whose 
great merit is that the exact ground-state energy and one-body density of the complicated 
many-body system is obtained from minimizing a suitable energy density 
functional (EDF). To make the problem tractable, Kohn and Sham demonstrated how 
the complex interacting many-body system can be made equivalent to a system of 
non-interacting electrons moving in an external---mean-field-like---potential\,\cite{Kohn:1965}. 
Among the advantages of the Kohn-Sham formulation is that self-consistent problems 
of this kind are routinely solved in many fields, including nuclear physics. Indeed, nuclear 
EDFs, although not always known as such, have a long and successful history in nuclear 
physics; see Ref.\,\cite{Furnstahl:2019lue} and references contained therein. The
widely used density-dependent Skyrme forces were developed almost a decade 
before the inception of density functional theory\,\cite{Skyrme:1956zz,Skyrme:1959zz}. 
In this paper we implement a covariant formulation of DFT that is based on an extension 
of the work by Walecka, Serot, and many others\,\cite{Serot:1984ey}. For details of the 
particular implementation  used in this work, see the recent review published in 
Ref.\cite{Yang:2019fvs}.

Advocating in favor of mean-field-like approaches for light systems such as ${}^{8}{\rm He}$ 
may come as a surprise. Although not a problem in the case of electrons bound to a heavy 
nucleus, the main problem with self-bound systems such as atomic nuclei is the absence of
a natural external potential and a proper treatment of the center of mass (COM). Indeed, 
as pointed out by Engel\,\cite{Engel:2006qu}, without a proper decoupling of the COM, the 
ground state of a self-bound system has a---manifestly incorrect---density that is uniformly 
distributed over space\,\cite{Engel:2006qu,Furnstahl:2019lue}. Among the 
treatments dealing with the removal of the COM contribution to the energy is an approach
based on a harmonic-oscillator approximation. This prescription, which falls down slowly 
with mass number\,\cite{Brown:1998}, makes a significant contribution to the energy of light 
nuclei---especially for those at the drip line. As such, large COM corrections to the energy 
hinder any meaningful prediction of the ground-state energy of light systems. However, the 
situation improves considerably when dealing with the linear response of the system. More 
than six decades ago in a seminal paper, Thouless showed how in a self-consistent formulation, 
the spurious state associated with a uniform translation of the center of mass separates out 
cleanly from the physical modes by having its strength shifted to zero excitation 
energy\,\cite{Thouless:1961}. This result is particular relevant for isoscalar dipole excitations 
that share the same quantum numbers as the center of mass. However, for neutron-rich 
nuclei such as ${}^{8}{\rm He}$, one expects a strong mixing between isoscalar and isovector 
dipole modes. It is the main goal of the present contribution to examine the impact of such a 
mixing on the emergence---or lack-thereof---of a soft dipole mode in ${}^{8}{\rm He}$.

The paper has been organized as follows. In Sec.\,\ref{Formalism} a brief description of the 
covariant RPA formalism used in this work is presented, paying special attention to the 
treatment of the continuum and the mixing between isoscalar and isovector modes. 
Self-consistent results are then presented in Sec.\,\ref{Results} for the ground-state 
properties and distribution of isovector dipole strength of ${}^{8}{\rm He}$. Finally, 
Sec.\,\ref{Conclusions} contains a summary of the main results.
\bigskip

\section{Formalism}
\label{Formalism}

The energy density functional used in this work is based on the non-linear model introduced in 
Ref.\,\cite{Mueller:1996pm}, supplemented by an isoscalar-isovector term that influences the 
dynamics of neutron-rich matter\,\cite{Horowitz:2000xj}. Although previously discussed in great 
detail elsewhere, see for example Ref.\,\cite{Chen:2014sca} and references contained therein, 
we display for completeness the interacting Lagrangian density:
\begin{widetext}
\begin{eqnarray}
 {\mathscr L}_{\rm int} &=&
\bar\psi \left[g_{\rm s}\phi   \!-\! 
         \left(g_{\rm v}V_\mu  \!+\!
    \frac{g_{\rho}}{2}{\mbox{\boldmath $\tau$}}\cdot{\bf b}_{\mu} 
                               \!+\!    
    \frac{e}{2}(1\!+\!\tau_{3})A_{\mu}\right)\gamma^{\mu}
         \right]\psi \nonumber \\
                   &-& 
    \frac{\kappa}{3!} (g_{\rm s}\phi)^3 \!-\!
    \frac{\lambda}{4!}(g_{\rm s}\phi)^4 \!+\!
    \frac{\zeta}{4!}   g_{\rm v}^4(V_{\mu}V^\mu)^2 +
   \Lambda_{\rm v}\Big(g_{\rho}^{2}\,{\bf b}_{\mu}\cdot{\bf b}^{\mu}\Big)
                           \Big(g_{\rm v}^{2}V_{\nu}V^{\nu}\Big)\;,
 \label{LDensity}
\end{eqnarray}
\end{widetext}
where the isodoublet nucleon field $\psi$ interacts through the exchange of photons 
($A_{\mu}$) and three ``mesons" of diverse spin-isospin character: a scalar-isoscalar 
($\phi$) a vector-isoscalar ($V^{\mu}$), and a vector-isovector 
(${\bf b}_{\mu}$)\,\cite{Mueller:1996pm}. Further, to improve the predictive power of 
the model, various self-interacting meson terms have been added. Ground-state 
properties of the system---namely, single-particle energies and Dirac orbitals, 
one-body densities, and mean-field-like potentials---are obtained from a
self-consistent solution of the Kohn-Sham equations\,\cite{Yang:2019fvs}. 

\begin{figure}[ht]
\vspace{-0.05in}
\includegraphics[width=0.48\textwidth]{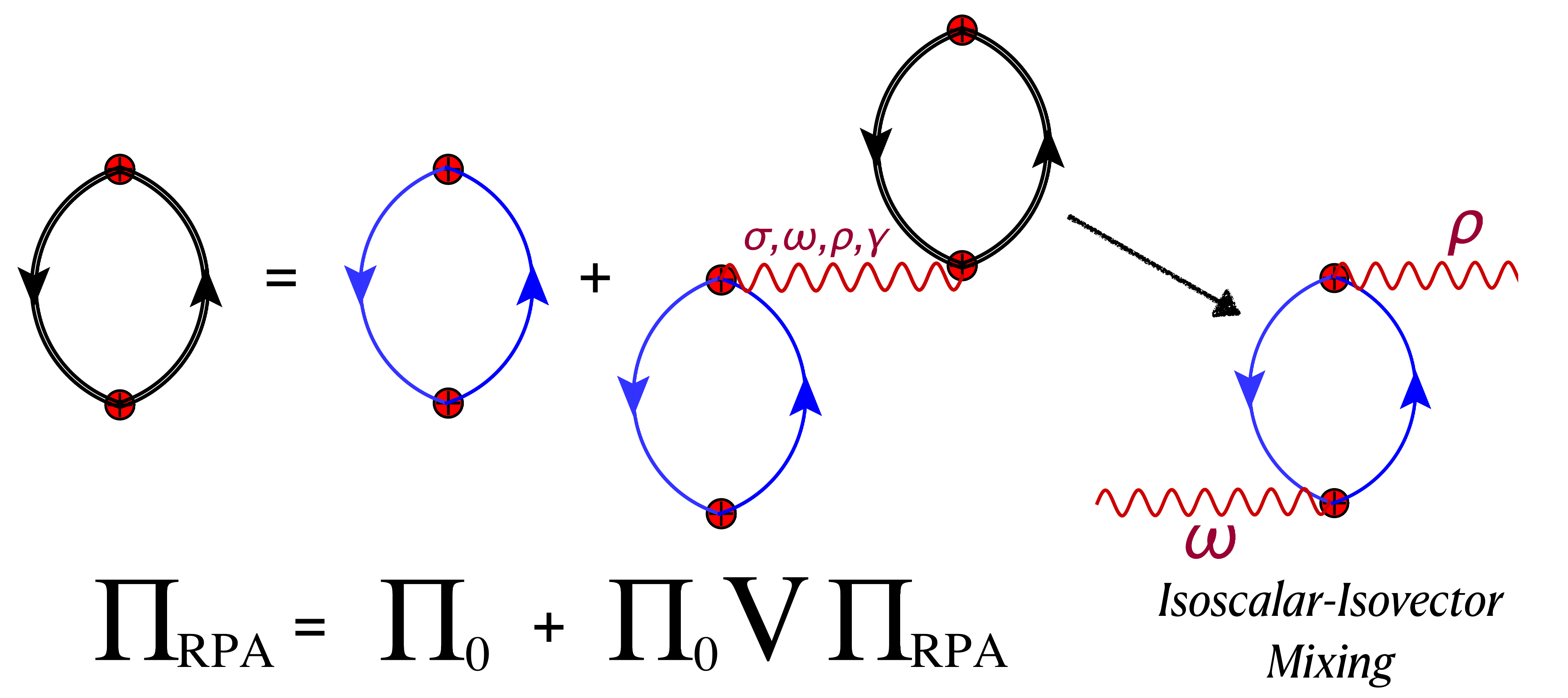}
\caption{(Color online) Diagrammatic representation of the RPA equations. 
The bubble with the thick lines represents the fully correlated polarization
tensor, while the one depicted with the thin  lines is the uncorrelated
polarization. The residual interaction denoted with the wavy line must
be identical to the one used to generate the ground-state. The arrow in 
the figure indicates that the the RPA bubble contains mixed contribution
of various isospin and Lorentz structures.}
\label{Fig1}
\end{figure}

Given that the Kohn-Sham equations may be derived from a variational approach, 
one can examine the small oscillations around the ground state. The consistent linear 
response of the ground state to an external perturbation is encapsulated in the RPA 
formalism that ensures that important symmetries are preserved\,\cite{Dickhoff:2005,
Piekarewicz:2013bea}. Particularly critical to this work is the decoupling of the spurious 
state associated with a uniform translation of the center of mass\,\cite{Thouless:1961}. 

The first step in generating the RPA response is the calculation of the uncorrelated
polarization tensor, depicted by the thin (blue) bubble in Fig.\,\ref{Fig1}. The spectral 
content of the uncorrelated polarization is both simple and illuminating: it contains 
simple poles at the single-particle excitations of the system with the associated 
transition densities obtained from the residues at the pole\,\cite{Dickhoff:2005}. 
One obtains the RPA polarization tensor, depicted by the thick (black) bubble in 
Fig.\,(\ref{Fig1}), by iterating the uncorrelated polarization to all orders. If many 
particle-hole pairs with the same quantum numbers are involved, then the RPA 
response is strongly collective and one ``giant resonance" tends to dominate, 
namely, the resonance exhausts most of the classical sum rule\,\cite{Harakeh:2001}. 

The diagrammatic structure of the RPA equations is depicted in Fig.\,\ref{Fig1}.
Two aspects of the RPA equations are particularly important. First, the wavy lines in
the figure denote the residual particle-hole interaction. It is only by using a residual 
particle-hole interaction consistent with the interaction used to generate the 
mean-field ground state that the spurious strength associated with a uniform 
translation of the center of mass is decoupled from the physical response. Second, 
the variety of isospin and Lorentz structures of the residual interaction leads to a 
highly-complex set of RPA equations. In particular, for nuclei with large neutron excess, 
the mixing of isoscalar and isovector modes is strong\,\cite{Piekarewicz:2013bea}. This
is illustrated by the arrow in the figure that indicates that the RPA bubble contains
mixed isoscalar-isovector contributions. It is precisely such strong isoscalar-isovector 
mixing that will become critical in our interpretation of the emergence, or lack-thereof, 
of a soft dipole mode in ${}^{8}$He. 

We conclude this section by relating the distribution of isovector dipole strength $R(\omega)$
to the photoabsorption cross section and by defining various moments of the distribution. As 
shown in Ref.\,\cite{Piekarewicz:2021jte}, $R(\omega)$ may be obtained from the dynamic
longitudinal response, which is a function of both the excitation energy $\omega$ and the 
momentum transfer. In turn, the product $\omega R(\omega)$ is directly proportional
to the photoabsorption cross section, namely,
\begin{equation}
 \sigma_{\!\rm abs}(\omega) = \frac{16\pi^{3}}{9}\frac{e^{2}}{\hbar c}
 \omega R(\omega).
\label{PhotoAbs}
\end{equation}
Often used in the literature are moments of the distribution of strength which are defined
as follows:
\begin{equation} 
  m_{n}=\int_{0}^{\infty} \!\!\omega^{n} R(\omega)\hspace{1pt}d\omega.
\label{Moments}
\end{equation}
In particular, the energy weighted sum $m_{1}$ satisfies a classical sum rule\,\cite{Harakeh:2001}, 
whereas the inverse energy weighted sum $m_{-1}$ is proportional to the electric dipole polarizability
\alphaD\,\cite{Roca-Maza:2013mla}---a physical observables that has been shown to be a good 
isovector indicator\,\cite{Reinhard:2010wz,Piekarewicz:2012pp}. That is,
\begin{subequations}
\begin{align}
 m_{1} & =
 \frac{9\hbar^{2}}{8\pi M}\left(\frac{NZ}{A}\right)\!\approx\!
 14.8 \left(\frac{NZ}{A}\right) {\rm fm}^{2}\,{\rm MeV} \;, 
 \label{EWSR}\\
 \text{\alphaD}  & = \frac{\hbar c}{2\pi^{2}} \int_{0}^{\infty} 
 \frac{\sigma_{\!\rm abs}(\omega)}{\omega^{2}}\,d\omega =
 \frac{8\pi e^2}{9} m_{-1}. 
\label{alphaD}
\end{align}
\end{subequations}

\section{Results}
\label{Results}

Following the organizational scheme of Ref.\cite{Bonaiti:2021kkp}, one starts 
this section by presenting results for the ground-state properties of ${}^{8}$He 
followed by a discussion on the distribution of dipole strength. Predictions are 
made using three covariant energy density functionals: RMF016 (also known 
as ``FSUGarnet"), RMF022, and RMF028 (or ``FSUGold2")\,\cite{Chen:2014mza}. 
All three EDFs are identical in the isoscalar sector but differ in their isovector 
properties. Specifically, the EDFs were calibrated assuming different values for 
the (at the time) unknown value of the neutron skin thickness of ${}^{208}$Pb.
In particular, RMF016 was calibrated assuming a neutron skin thickness of 
0.16\,fm, RMF022 of 0.22\,fm, and RMF028 of 0.28\,fm. Based on the result 
published by the PREX collaboration\,\cite{Adhikari:2021phr}, namely,
$R_{\rm skin}^{208}\!=\!0.283\pm0.071\,{\rm fm}$, the RMF016 prediction 
falls within the two-sigma interval. 

One should note that within the context of covariant
DFT, all three accurately calibrated EDFs have been successful in describing
a host of physical observables, such as ground-state properties of 
medium- to heavy-mass nuclei, their linear response, and the structure of
neutron stars. Moreover, such EDFs have also been used to explore the 
evolution of the ground-state energy of the oxygen isotopes\,\cite{Chen:2014mza}. 
Whereas no lighter system than oxygen has been studied with this set of EDFs,
it is interesting to explore their predictions for the isovector dipole response of 
${}^{8}$He, primarily due to the critical role that self-consistency plays in 
eliminating any spurious contamination.

\subsection{Ground State Properties}
\label{gsp}

\begin{figure}[ht]
\vspace{-0.1in}
\includegraphics[width=0.45\textwidth]{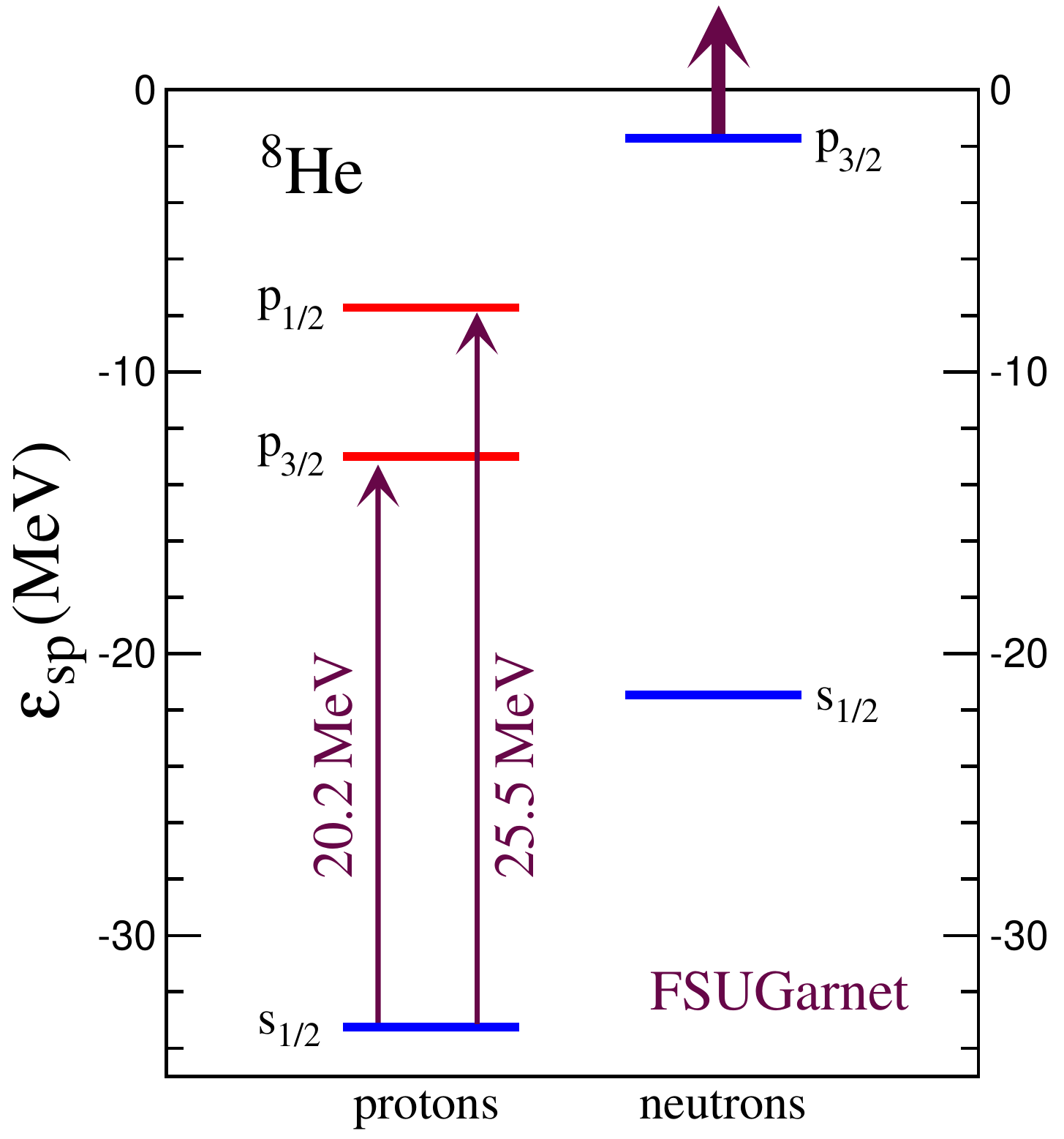}
\caption{(Color online) Single-particle spectrum for ${}^{8}$He as 
predicted by the covariant energy density functional FSUGarnet=RMF016.
The blue(red) lines denote occupied(empty) orbitals and the 
thin arrows indicate discrete excitations into bound 
states. In turn, the thick arrow indicates that low-energy strength is
expected to emerge from the excitation of the weakly-bound neutron 
$p_{3/2}$ orbital into the continuum.}
\label{Fig2}
\end{figure}

Self-consistent predictions for the bound single-particle spectrum of ${}^{8}$He 
are displayed in Fig.\ref{Fig2}, with the blue(red) lines indicating the occupied(vacant)
single-particle orbitals. The two thin arrows indicate the two lowest ``sharp" dipole 
transitions on the proton side. In contrast, all dipole excitations on the neutron side 
involve transitions into the continuum. Within the DFT framework employed here, 
the soft-dipole excitations indicated by the thick arrow involve the transition of the
weakly bound $p_{3/2}$ orbital into the $sd$ shell, which lies entirely in the 
continuum. These low-energy excitations will be discussed in greater detail in 
Sec.\ref{dr}.

\begin{figure}[ht]
\vspace{-0.05in}
\includegraphics[width=0.4\textwidth]{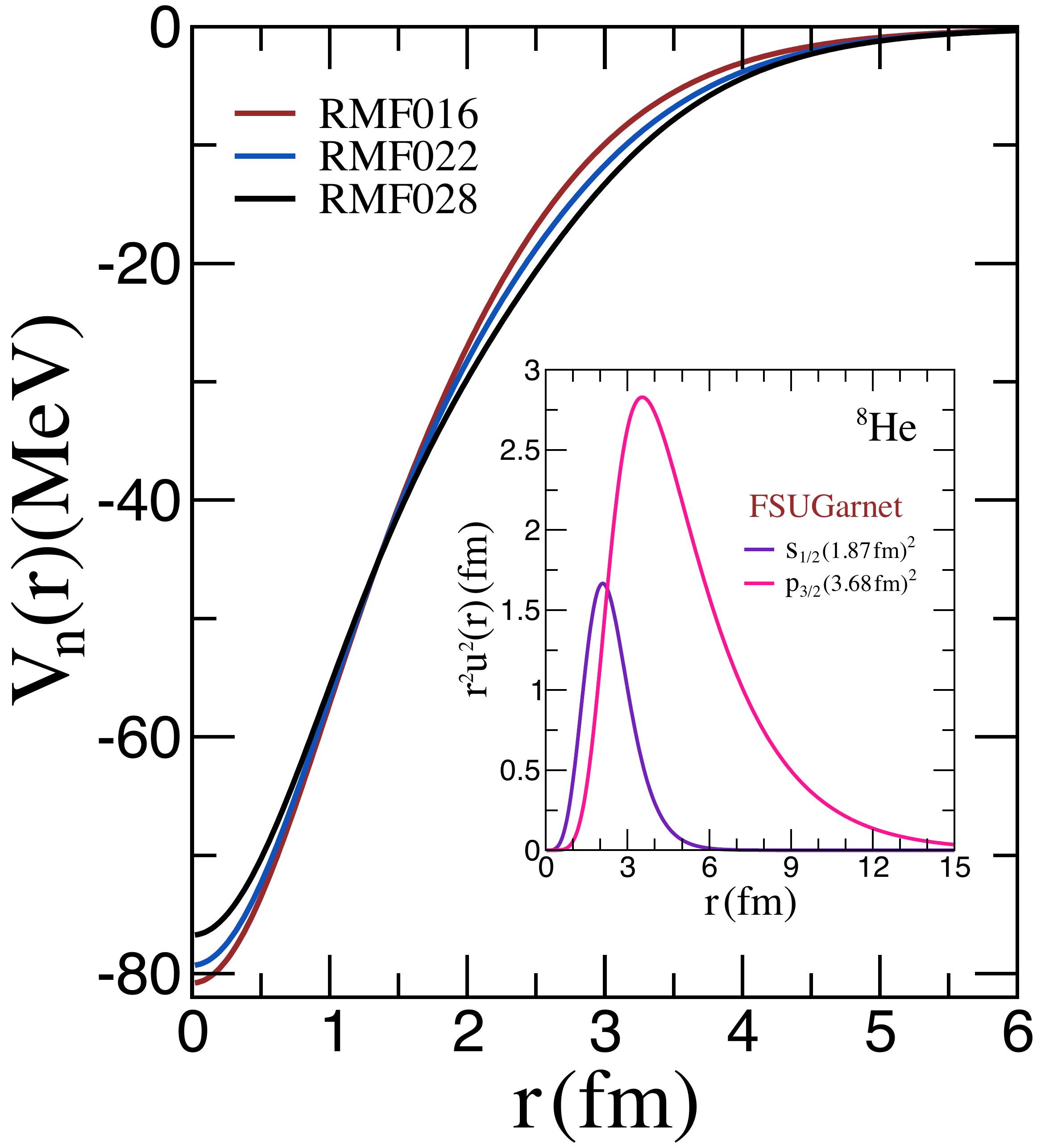}
\caption{(Color online) Effective ``Schr\"odinger-like" neutron 
potential for the three models considered in this work. The 
inset shows the two bound neutron orbitals in ${}^{8}$He 
supported by the RMF016=FSUGarnet potential and 
illustrates the large spatial extent of the $p_{3/2}$ orbital.}
\label{Fig3}
\end{figure}

Listed in Table\,\ref{Table1} are energies and root-mean-square radii for 
${}^{8}$He as predicted by the three models introduced earlier. The second 
column lists the single-particle energy of the $p_{3/2}$ neutron orbital
which displays a significant model dependence that is attributed to the 
difference in the isovector properties of the models. As shown in Fig.\ref{Fig3}, 
the model with the stiffest symmetry energy (RMF028) generates the most 
attractive neutron potential at the large distances of relevance to the 
weakly-bound $p_{3/2}$ orbital. Indeed, as indicated in the inset to 
Fig.\ref{Fig3}, the $p_{3/2}$ orbital peaks at a distance of about 3.7\,fm 
where the neutron potential generated by the RMF028 model is about 
2 MeV deeper than the one generated by the model with the softest 
symmetry energy (RMF016). Note that the neutron potential is 
an effective Schr\"odinger-like potential obtained from a linear combination 
of the relativistic scalar and vector potentials\,\cite{Serot:1984ey}. 

\begin{widetext}
\begin{center}
\begin{table}[h]
\begin{tabular}{|l||c|c||c|c|c||c|c|}
 \hline\rule{0pt}{2.25ex}   
 \!\!Model & $\varepsilon(p_{3/2})$(MeV) & $E/A$(MeV) & $R_{p}$(fm) 
                & $R_{n}$(fm) & $R_{n}\!-\!R_{p}$(fm) & $R_{\rm ch}$(fm) & $R_{\rm wk}$(fm) \\
 \hline
 \hline\rule{0pt}{2.25ex} 
 \!\!RMF016   &  1.714  & 2.241-3.764 & 1.897   &  3.206   &  1.309 & 1.998 & 3.354 \\ 
     RMF022   &  2.740  & 2.521-4.044 & 1.883   &  3.023   &  1.140 & 1.981 & 3.175  \\
     RMF028   &  3.784  & 2.785-4.308 & 1.876    &  2.904   &  1.028 & 1.970 & 3.060  \\
 \hline\rule{0pt}{2.25ex} 
 Experiment   &   2.535(8)     & 3.925  & 1.807(28)  &  2.73(9) &  0.92(10) & 1.929(26) & ---  \\
 \hline                                                                                                 
\end{tabular}
\caption{Predictions for a few ground-state properties of ${}^{8}$He for the three
models used in this work. The binding energy of the neutron $p_{3/2}$ orbital is
compared against the experimental one-neutron separation energy listed in the
National Nuclear Data Center database. The quoted experimental energy per nucleon 
was obtained from Refs.\,\cite{Brodeur:2011sam,Huang:2021nwk,Wang:2021xhn}, 
the experimental charge radius from Ref.\,\cite{Mueller:2007dhq}, while the derived 
quantities for $R_{p}$ and $R_{n}$ were extracted from Ref.\,\cite{Liu:2021cbn}.}
\label{Table1}
\end{table}
\end{center}
\end{widetext}

The third column in Table\,\ref{Table1} displays the binding energy per nucleon 
and makes abundantly clear one of the problems of using DFT for a light, 
self-bound system such as ${}^{8}$He. The lower value listed on the table
does not include any center-of-mass correction, while the higher value includes 
a significant COM correction of 1.52\,MeV, obtained by assuming a harmonic 
oscillator approximation\,\cite{Brown:1998}. Note that the lightest nucleus that
was used in the calibration of the three covariant EDFs was 
${}^{16}$O\,\cite{Chen:2014mza,Chen:2014sca}, twice as heavy as ${}^{8}$He.

The rest of the columns in Table\,\ref{Table1} are predictions for rms radii. Based 
on the statistical analysis carried out in Ref.\cite{Chen:2014sca}, an error of at 
least 0.03\,fm should be attached to all theoretical predictions. Although several 
``experimental" values are listed in the table, only the charge radius of ${}^{8}$He 
can be regarded as a model-independent 
determination\,\cite{Mueller:2007dhq,Brodeur:2011sam,Krauth:2021foz}. 
Instead, the proton radius $R_{p}$ quoted in Table\,\ref{Table1} requires the
unfolding of the finite proton size\,\cite{Liu:2021cbn}. However, as indicated in 
Eq.(19) of Ref.\,\cite{Horowitz:2012we}, the charge radius includes spin-orbit 
contributions that go above and beyond the finite size of the proton. In the
case of the experimental neutron radius quoted in Table\,\ref{Table1}, it was 
obtained from both $R_{p}$ and a determination of the matter radius from 
an elastic proton scattering experiment\,\cite{Liu:2021cbn}.  However, besides 
the inherent uncertainties involved in the determination of nuclear radii using 
hadronic probes\,\cite{Thiel:2019tkm}, the determination of $R_{n}$ is also 
hindered by the uncertainties in the extraction of $R_{p}$ mentioned above. 

\begin{figure}[ht]
\vspace{-0.05in}
\includegraphics[width=0.48\textwidth]{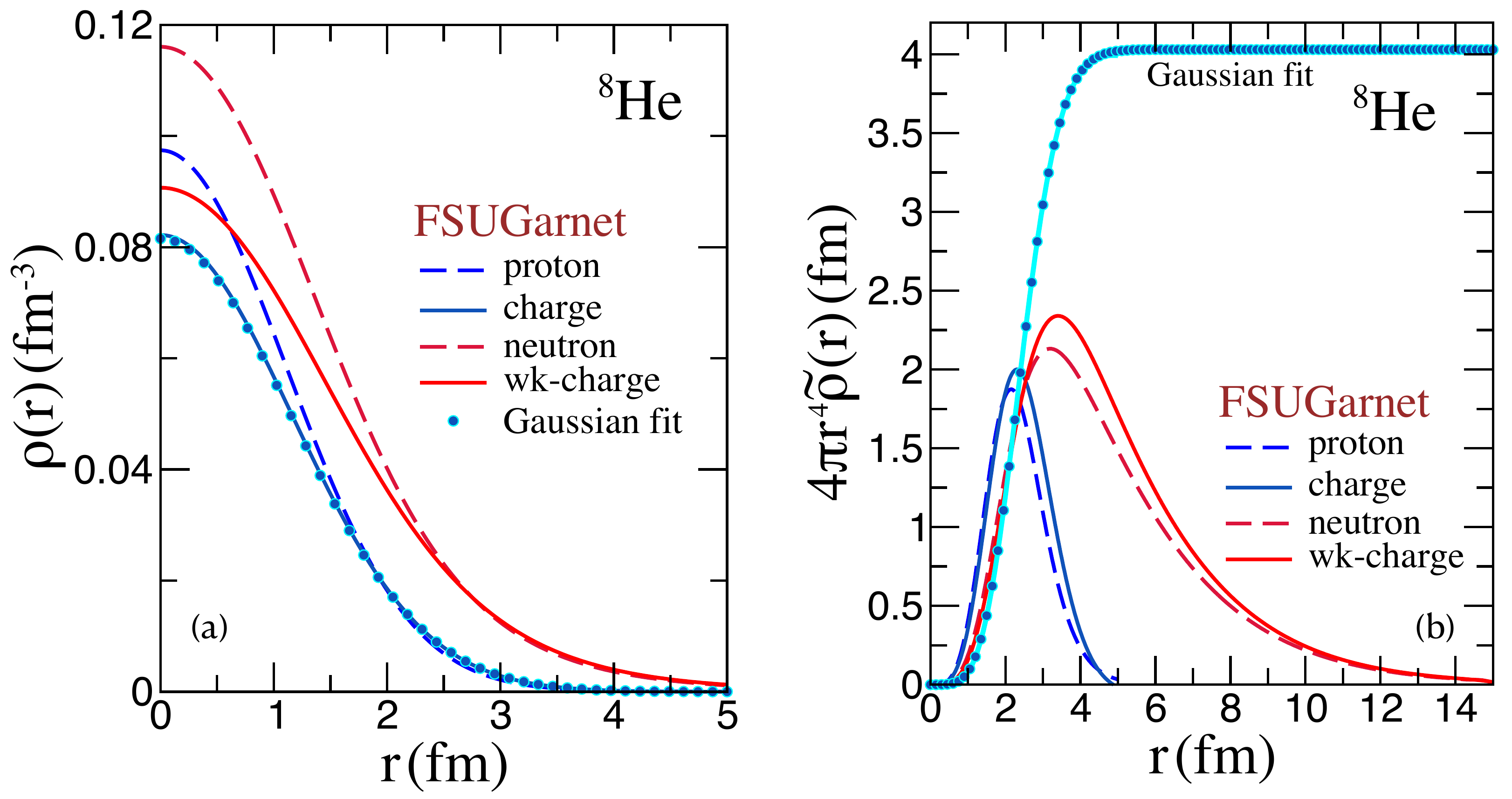}
\caption{(Color online) (a) proton, neutron, charge, and weak-charge
densities for ${}^{8}$He as predicted by the relativistic FSUGarnet 
density functional. The dots represent a one-parameter Gaussian fit 
to the charge density. (b) Ground-state densities suitably scaled so
that the area under the curve equals the mean-square radius of the
distribution.}
\label{Fig4}
\end{figure}

Within the context of density functional theory and the Kohn-Sham equations, 
one has access to the entire spatial distributions, from which radii---as well as 
any other moment of the distribution---may be computed. Proton, neutron, charge, 
and weak-charge densities are displayed in Fig.\,\ref{Fig4}(a) as predicted by  
FSUGarnet=RMF016. Note that both the charge and weak-charge densities 
incorporate spin-orbit corrections as outlined in Ref.\,\cite{Horowitz:2012we}. 
In all four cases the spatial distribution can be accurately fitted by a 
one-parameter Gaussian form. For example, in the case of the charge 
density and its associated form factor one obtains

\begin{subequations}
\begin{align}
    \rho_{\rm ch}(r) & =\left( \frac{3Z}{2\pi R_{\rm ch}^{2}}\right)^{\!\!3/2}
     \!\!\!\mathlarger{e}^{-3r^{2}/2R_{\rm ch}^{2}},  \\ 
     F_{\rm ch}(q) & = \mathlarger{e}^{-q^{2}R_{\rm ch}^{2}/6}, 
 \label{Gaussian}
\end{align}
\end{subequations}
where $Z$ is the nuclear charge, $R_{\rm ch}$ is the charge radius of the distribution, 
and the form factor has been normalized to $F_{\rm ch}(q\!=\!0)\!=\!1$. 
The Gaussian fit to the charge density is displayed with the small circles in Fig.\,\ref{Fig4}(a). 
Plotted in Fig.\,\ref{Fig4}(b) is a quantity for which the area under the curve equals the 
mean-square radius. The circles in the figure denotes the cumulative (or running) sum of 
the charge distribution and converges to $R_{\rm ch}^{2}\!\approx\!(2\,{\rm fm})^{2}$. 

So what can be concluded from comparing the experimental results against a theoretical 
framework that is likely being pushed beyond its limits of applicability. Insofar as the energy 
per nucleon is concerned, the violation of translation symmetry inherent to any 
mean-field-like description results in a center-of-mass correction that makes a significant 
contribution to the total energy of the system, calling into question the relevance of the
predictions. However, COM corrections to the charge radius are relatively 
small\,\cite{Mihaila:1998qr} and comparable to the statistical error obtained in the calibration 
of the functional. Further, the experimental value quoted in Table\,\ref{Table1} is only one of 
three experimental determinations of the charge radius of ${}^{8}$He. Taking into account all 
the measurements up to date\,\cite{Mueller:2007dhq,Brodeur:2011sam,Krauth:2021foz}, one 
obtains at the one-sigma level an estimate of the charge radius of ${}^{8}$He that lies in the 
interval $1.903\,\lesssim\,R_{\rm ch}(\rm fm)\,\lesssim1.975$. This, together with the 0.03\,fm
theoretical uncertainty, yields a prediction for the charge radius that appears to be in reasonably 
good agreement with experiment. Finally,  an inescapable consequence of the small 
one-neutron separation energy is the emergence of low-energy dipole strength in the 
uncorrelated (single-particle) response. How the dipole strength rearranges as a 
result of the inclusion of RPA correlations is the main topic of the next section.

\subsection{Dipole Response}
\label{dr}

In the previous section several ground state properties of ${}^{8}$He were discussed. 
As alluded earlier, a self-consistent solution to the Kohn-Sham equations yield: 
(a) single-particle energies and Dirac orbitals, (b) ground-state densities, and (c) the  
self-consistently determined mean-field (or Kohn-Sham) potential. Critical to the 
consistency of the formalism is that the potential so determined, must be used without 
modification to generate the single-nucleon propagator from which the uncorrelated 
polarization tensor is obtained\,\cite{Piekarewicz:2001nm,Piekarewicz:2013bea}. 
Moreover, to avoid any reliance on artificial cutoffs and truncations, the nucleon propagator 
(depicted by the thin line in Fig.\,\ref{Fig1}) is computed non-spectrally using Green's 
function methods\,\cite{Dickhoff:2005}.  

\begin{figure}[ht]
\vspace{-0.05in}
\includegraphics[width=0.40\textwidth]{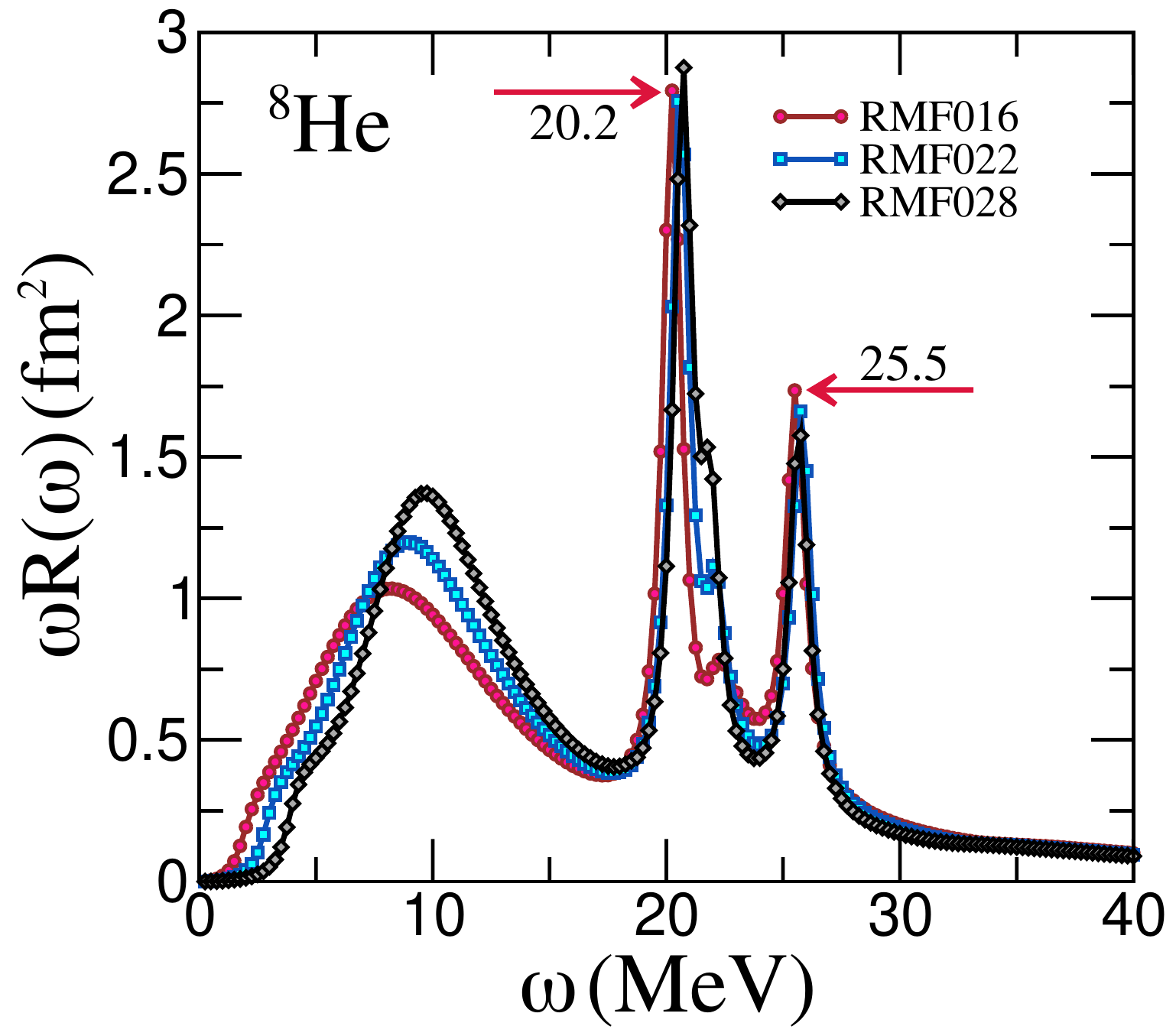}
\caption{(Color online) Uncorrelated energy-weighted dipole response for ${}^{8}$He
for the three models considered in the text. The uncorrelated response is made up
of individual particle-hole excitations with the correct quantum numbers. The arrows
indicate the location of the proton excitations based on the single-particle spectrum
displayed in Fig.\ref{Fig2}.}
\label{Fig5}
\end{figure}

The uncorrelated dipole response $R(\omega)$, weighted by the excitation energy 
$\omega$, is displayed in Fig.\ref{Fig5}. Clearly visible in the figure are the two sharp 
proton transitions involving the excitation of the $s_{1/2}$ orbital into the bound 
$p_{3/2}$-$p_{1/2}$ spin-orbit partners, in perfect agreement with the single-particle 
spectrum displayed in Fig.\ref{Fig2}. Also shown in the figure is the emergence of
low-energy dipole strength resulting from the excitation of the $p_{3/2}$ neutron orbital 
into the continuum. Note that among the advantages of displaying the energy weighted 
dipole response is that the area under the curve is directly related to ``classical" energy 
weighted sum rule (EWSR) given in Eq.(\ref{EWSR})\,\cite{Harakeh:2001}. That is,
\begin{equation} 
  m_{1} \approx 14.8 \left(\frac{NZ}{A}\right){\rm MeV\,fm^{2}} 
   \xrightarrow[\text{}]{\text{${}^{8}$He}} 22.2\,{\rm MeV\,fm^{2}}.
 \label{EWSRHe8}
\end{equation}
For the uncorrelated response displayed in Fig.\ref{Fig5}, the energy weighted sum is 
predicted to be equal to 22.4, 22.6, and 22.8$\,{\rm MeV\,fm^{2}}$ for RMF016, RMF022, 
and RMF028, respectively---in excellent agreement with the classical EWSR. 

One now proceeds to discuss the RPA response, which represents the consistent linear 
response of the ground state to an external perturbation\,\cite{Dickhoff:2005}. As depicted 
in Fig.\ref{Fig1}, the RPA response goes beyond the single-particle response by building 
collectivity through the coherent contribution of many particle-hole pairs. Although large
center-of-mass corrections preclude meaningful prediction of the ground state energy of 
${}^{8}$He, the self-consistent RPA response offers a unique and powerful solution to the 
center-of-mass problem: spurious states associated with a uniform translation of the 
center of mass decouple from the physical modes by having their strength shifted to zero 
excitation energy\,\cite{Thouless:1961}. This is particularly relevant to the distribution of
isoscalar dipole ($J^{\pi}\!=\!1^{-}, T\!=\!0$) strength that shares the same quantum numbers 
as the center of mass. But given that for nuclei with a significant neutron excess, such as 
${}^{8}$He, a significant mixing between the isoscalar and isovector modes is expected, 
the possible emergence of a soft dipole mode will undoubtedly be affected by the decoupling 
of the spurious center-of-mass mode.

\begin{figure}[ht]
\vspace{-0.05in}
\includegraphics[width=0.49\textwidth]{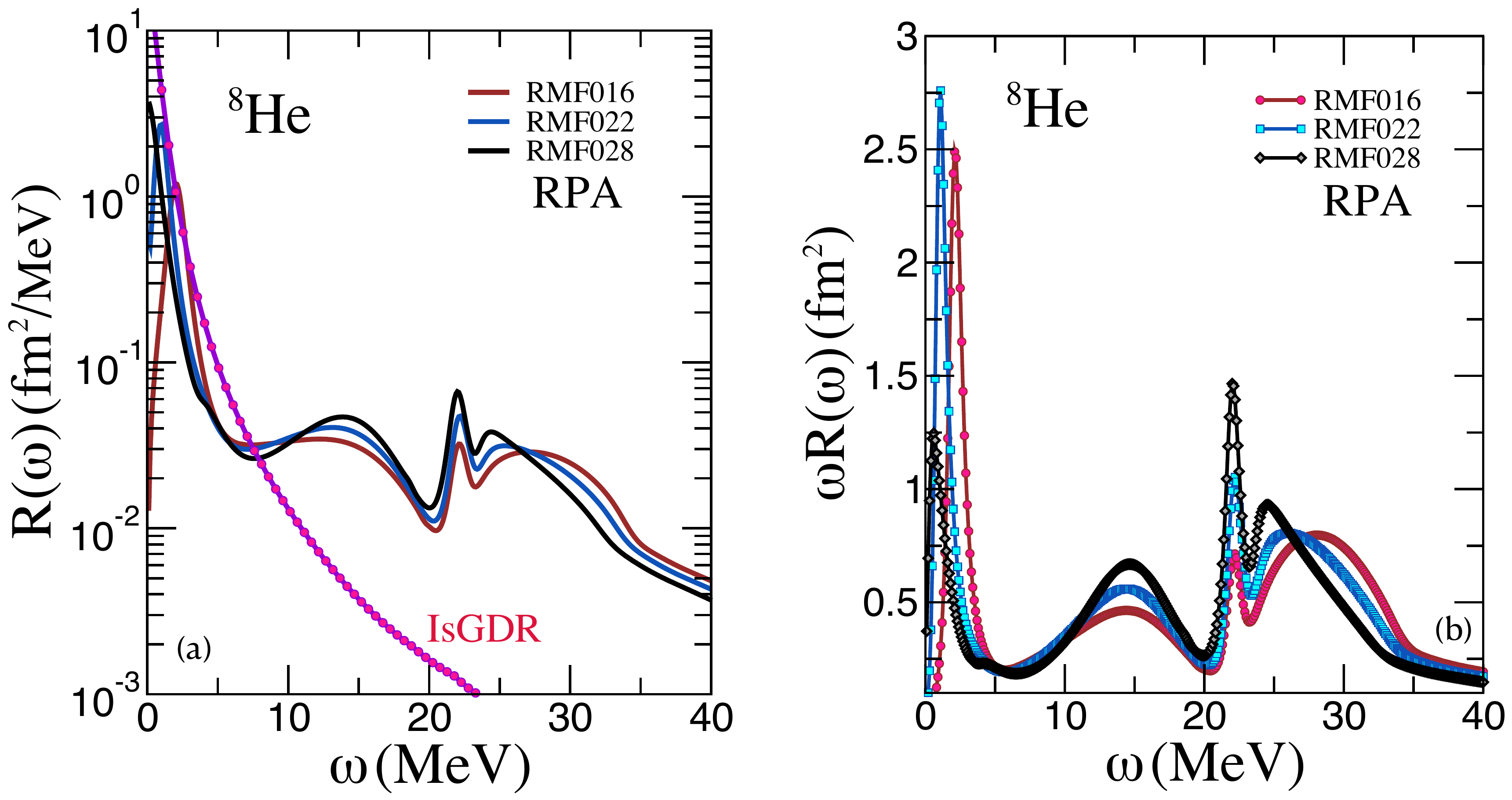}
\caption{(Color online) (a) Correlated (RPA) dipole response for ${}^{8}$He for the three 
models considered in the text. Also shown is the isoscalar dipole response to illustrate
the migration of the spurious mode to zero excitation energy. (b) The narrow structures 
appearing at low energies in the energy weighted RPA response are associated with the 
spurious center-of-mass mode.}
\label{Fig6}
\end{figure}

To investigate the mixing between modes, the distribution of isovector dipole strength obtained
from a self-consistent covariant RPA calculation is displayed (on a logarithmic scale) on 
Fig.\,\ref{Fig6}(a). Also shown is the distribution of isoscalar dipole strength predicted by the 
RMF028 model. As argued by Thouless\,\cite{Thouless:1961}, the spurious state associated 
with the translation of the center of mass is shifted to zero excitation energy. Indeed it appears 
that most (if not all!) of the uncorrelated isoscalar dipole strength shown in Fig.\,\ref{Fig5} is 
shifted to zero energy; note that the uncorrelated response is identical in both the isoscalar 
and isovector channels. Given the anticipated strong mixing between the isoscalar and isovector 
dipole modes, it is reasonable to identify the narrow structures appearing at low energies in the 
isovector dipole response---best seen in Fig.\,\ref{Fig6}(b)---as contaminants associated with 
the spurious center-of-mass mode. Thus, the theoretical formalism implemented here disfavors 
the emergence of a soft dipole mode in ${}^{8}$He---in agreement with the conclusions from 
Refs.\cite{Iwata:2000dd,Holl:2021bxg}.

\begin{figure}[ht]
\vspace{-0.05in}
\includegraphics[width=0.45\textwidth]{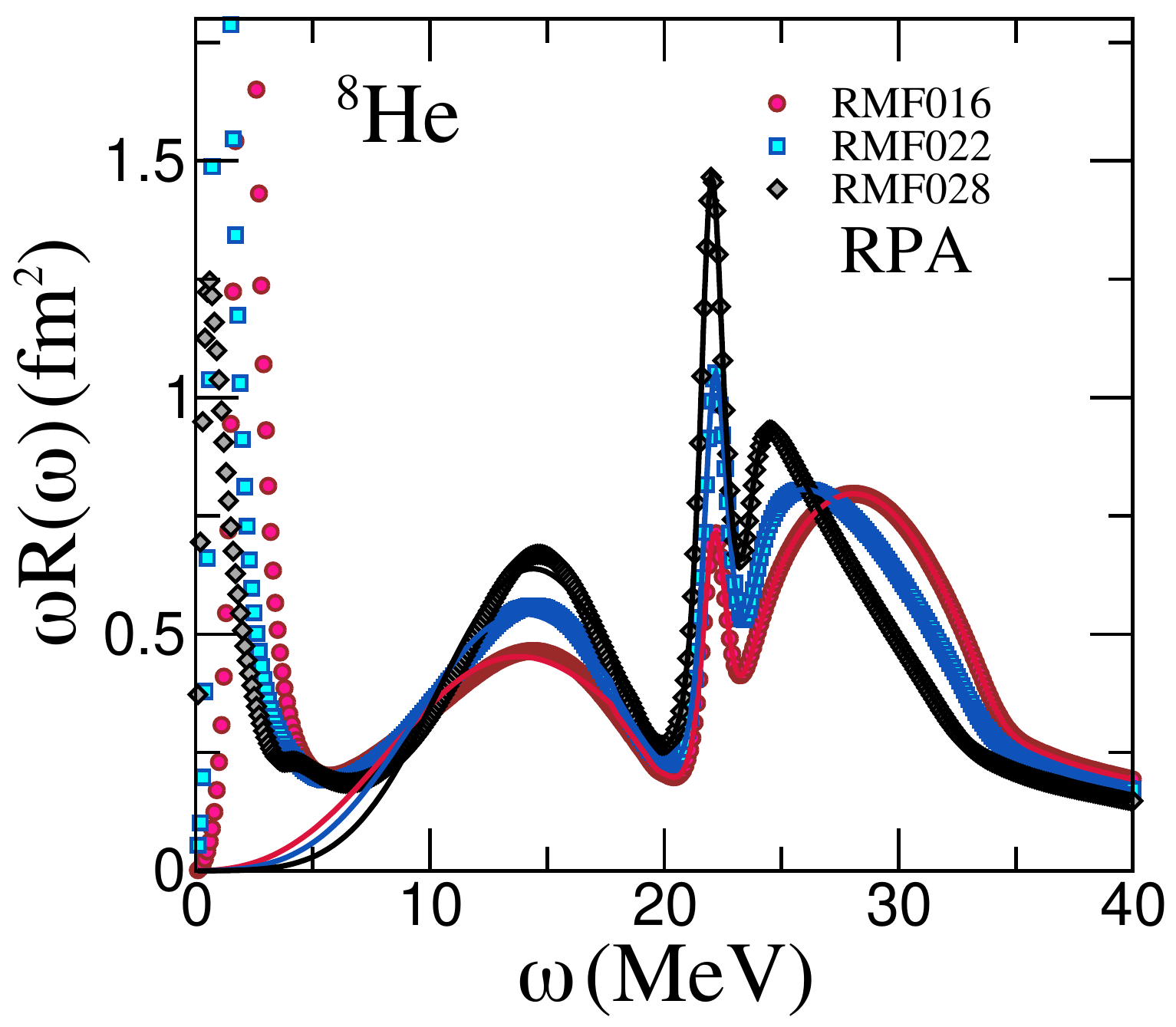}
\caption{(Color online) Correlated (RPA) energy-weighted dipole response for ${}^{8}$He 
as predicted for the three models considered in the text. The solid lines at low excitation 
energy represent an ad-hoc attempt to remove the spurious strength in favor of a smooth 
extrapolation to zero excitation energy.}
\label{Fig7}
\end{figure}

In an effort to remove the spurious contribution from the isovector dipole response, a smooth 
extrapolation to zero frequency is implemented in Fig.\,\ref{Fig7}. By doing so, one can now 
provide estimates for the various moments of the distribution as listed in Table\,\ref{Table2}. 
\begin{table}[h]
\begin{tabular}{|l||c|c|c|c|}
 \hline\rule{0pt}{2.25ex}   
 \!\!Model & $m_{1}({\rm fm^{2}\,MeV})$ & $m_{0}({\rm fm^{2}})$ & 
                   $m_{-1}({\rm fm^{2}/MeV})$ & \alphaD$({\rm fm^{3}})$ \\
 \hline
 \hline\rule{0pt}{2.25ex} 
 \!\!RMF016   &  16.37  & 0.829 & 0.065   &  0.262   \\ 
     RMF022   &  16.70  & 0.849 & 0.060   &  0.242  \\
     RMF028   &  16.84  & 0.852 & 0.055   &  0.220   \\
 \hline                                                                                                 
\end{tabular}
\caption{Estimates for various moments of the isovector dipole response of  ${}^{8}$He, as
defined in Eq.(\ref{Moments}). Also shown is the electric dipole polarizability \alphaD. All 
these estimates are based on the smooth extrapolation to zero excitation energy depicted
in  Fig.\ref{Fig7}.} 
\label{Table2}
\end{table}

One should underscore that the estimates listed in Table\,\ref{Table2} are based on the 
removal of the spurious strength in favor of a smooth extrapolation to zero excitation energy. 
This largely ad-hoc procedure has a particularly strong effect on the electric dipole polarizability, 
which is particularly sensitive to the low-energy part of the response because of the 
$\omega^{-1}$ weighting. In the case of the energy weighted sum, the estimates are now 
significantly reduced relative to the classical EWSR quoted in Eq.(\ref{EWSR}). Finally, although 
the information encapsulated in the various moments is valuable, there is no substitute for 
a direct comparison between theory and experiment of the entire dipole distribution.

\section{Conclusions}
\label{Conclusions}

The fascinating dynamics of exotic neutron-rich nuclei has led to 
a paradigm shift in nuclear structure. Besides providing unique
insights into the limits of nuclear existence and the production of
heavy elements in the cosmos, the study of nuclei with large isospin
asymmetries offers meaningful experimental constraints on the
isovector sector of the nuclear interaction. In this paper the possible 
emergence of low-energy dipole strength in ${}^{8}$He was investigated,
a drip-line nucleus with the largest neutron-to-proton ratio known to
date. 

The possible existence of a soft dipole mode in ${}^{8}$He has been a 
highly controversial issue, with some experiments identifying low dipole 
strength at an excitation energy of about 3-4 MeV\,\cite{Markenroth:2001oxa,
Meister:2002uuu,Golovkov:2008kg,Grigorenko:2009} and others
refuting those claims\,\cite{Iwata:2000dd,Holl:2021bxg}. From the
theoretical perspective, a recent ab initio approach that merges the 
Lorentz integral transform with coupled-cluster theory reports a 
dipole response that shows strength at about 5 MeV. 

In this contribution, a theoretical formalism based on covariant 
density functional theory was used to examine the emergence of 
low energy dipole strength. Admittedly, using such a formalism 
for the study of a nucleus as light as ${}^{8}$He is questionable. 
Indeed, given that center-of-mass corrections fall down slowly 
with mass number, they make an appreciable contribution to 
the total energy of ${}^{8}$He, limiting the value of most
theoretical predictions.
However, the strength of DFT lies in its self-consistency. Whereas 
COM corrections to the ground-state energy may be large, any
spurious contamination from the COM is guaranteed to decouple 
from the physical isoscalar dipole response\,\cite{Thouless:1961}. 
This has important consequences for the isovector dipole response
because the mixing between the isoscalar and isovector modes is
anticipated to be strong for neutron-rich systems like ${}^{8}$He. 
Hence, the narrow structures that emerged at low energies in the 
isovector dipole response were attributed to the shift of the 
spurious strength to zero---or close to zero---excitation energy. 
Based on this interpretation, one concludes that the emergence 
of a soft dipole mode in ${}^{8}$He is disfavored by the adopted 
theoretical framework.

\begin{acknowledgments}
The author thanks Professor Kirby Kemper for many enlightening discussions
on the many experimental efforts devoted to understand the soft dipole response 
of ${}^{8}$He. This material is based upon work supported by the U.S. Department 
of Energy Office of Science, Office of Nuclear Physics under Award 
DE-FG02-92ER40750. 
\end{acknowledgments}

\bibliography{./He8IvGDR.bbl}

\begin{thebibliography}{48}%
\makeatletter
\providecommand \@ifxundefined [1]{%
 \@ifx{#1\undefined}
}%
\providecommand \@ifnum [1]{%
 \ifnum #1\expandafter \@firstoftwo
 \else \expandafter \@secondoftwo
 \fi
}%
\providecommand \@ifx [1]{%
 \ifx #1\expandafter \@firstoftwo
 \else \expandafter \@secondoftwo
 \fi
}%
\providecommand \natexlab [1]{#1}%
\providecommand \enquote  [1]{``#1''}%
\providecommand \bibnamefont  [1]{#1}%
\providecommand \bibfnamefont [1]{#1}%
\providecommand \citenamefont [1]{#1}%
\providecommand \href@noop [0]{\@secondoftwo}%
\providecommand \href [0]{\begingroup \@sanitize@url \@href}%
\providecommand \@href[1]{\@@startlink{#1}\@@href}%
\providecommand \@@href[1]{\endgroup#1\@@endlink}%
\providecommand \@sanitize@url [0]{\catcode `\\12\catcode `\$12\catcode
  `\&12\catcode `\#12\catcode `\^12\catcode `\_12\catcode `\%12\relax}%
\providecommand \@@startlink[1]{}%
\providecommand \@@endlink[0]{}%
\providecommand \url  [0]{\begingroup\@sanitize@url \@url }%
\providecommand \@url [1]{\endgroup\@href {#1}{\urlprefix }}%
\providecommand \urlprefix  [0]{URL }%
\providecommand \Eprint [0]{\href }%
\providecommand \doibase [0]{http://dx.doi.org/}%
\providecommand \selectlanguage [0]{\@gobble}%
\providecommand \bibinfo  [0]{\@secondoftwo}%
\providecommand \bibfield  [0]{\@secondoftwo}%
\providecommand \translation [1]{[#1]}%
\providecommand \BibitemOpen [0]{}%
\providecommand \bibitemStop [0]{}%
\providecommand \bibitemNoStop [0]{.\EOS\space}%
\providecommand \EOS [0]{\spacefactor3000\relax}%
\providecommand \BibitemShut  [1]{\csname bibitem#1\endcsname}%
\let\auto@bib@innerbib\@empty
\bibitem [{\citenamefont {Geesaman}\ \emph {et~al.}(2015)\citenamefont
  {Geesaman} \emph {et~al.}}]{LongRangePlan}%
  \BibitemOpen
  \bibfield  {author} {\bibinfo {author} {\bibfnamefont {D.}~\bibnamefont
  {Geesaman}} \emph {et~al.},\ }\enquote {\bibinfo {title} {Reaching for the
  horizon; the 2015 long range plan for nuclear science},}\ \ (\bibinfo {year}
  {2015})\BibitemShut {NoStop}%
\bibitem [{\citenamefont {Ahn}\ \emph {et~al.}(2019)\citenamefont {Ahn} \emph
  {et~al.}}]{Ahn:2019xgh}%
  \BibitemOpen
  \bibfield  {author} {\bibinfo {author} {\bibfnamefont {D.~S.}\ \bibnamefont
  {Ahn}} \emph {et~al.},\ }\href {\doibase 10.1103/PhysRevLett.123.212501}
  {\bibfield  {journal} {\bibinfo  {journal} {Phys. Rev. Lett.}\ }\textbf
  {\bibinfo {volume} {123}},\ \bibinfo {pages} {212501} (\bibinfo {year}
  {2019})}\BibitemShut {NoStop}%
\bibitem [{\citenamefont {Guillemaud-Mueller}\ \emph
  {et~al.}(1990)\citenamefont {Guillemaud-Mueller} \emph
  {et~al.}}]{Guillemaud-Mueller:1990vpa}%
  \BibitemOpen
  \bibfield  {author} {\bibinfo {author} {\bibfnamefont {D.}~\bibnamefont
  {Guillemaud-Mueller}} \emph {et~al.},\ }\href {\doibase
  10.1103/PhysRevC.41.937} {\bibfield  {journal} {\bibinfo  {journal} {Phys.
  Rev. C}\ }\textbf {\bibinfo {volume} {41}},\ \bibinfo {pages} {937} (\bibinfo
  {year} {1990})}\BibitemShut {NoStop}%
\bibitem [{\citenamefont {Hoffman}\ \emph {et~al.}(2009)\citenamefont {Hoffman}
  \emph {et~al.}}]{Hoffman:2009zza}%
  \BibitemOpen
  \bibfield  {author} {\bibinfo {author} {\bibfnamefont {C.}~\bibnamefont
  {Hoffman}} \emph {et~al.},\ }\href@noop {} {\bibfield  {journal} {\bibinfo
  {journal} {Phys. Lett.}\ }\textbf {\bibinfo {volume} {B672}},\ \bibinfo
  {pages} {17} (\bibinfo {year} {2009})}\BibitemShut {NoStop}%
\bibitem [{\citenamefont {Thoennessen}(2004)}]{Thoennessen:2004}%
  \BibitemOpen
  \bibfield  {author} {\bibinfo {author} {\bibfnamefont {M.}~\bibnamefont
  {Thoennessen}},\ }\href@noop {} {\bibfield  {journal} {\bibinfo  {journal}
  {Rep. Prog. Phys.}\ }\textbf {\bibinfo {volume} {67}},\ \bibinfo {pages}
  {1187} (\bibinfo {year} {2004})}\BibitemShut {NoStop}%
\bibitem [{\citenamefont {Markenroth}\ \emph {et~al.}(2001)\citenamefont
  {Markenroth} \emph {et~al.}}]{Markenroth:2001oxa}%
  \BibitemOpen
  \bibfield  {author} {\bibinfo {author} {\bibfnamefont {K.}~\bibnamefont
  {Markenroth}} \emph {et~al.},\ }\href {\doibase
  10.1016/S0375-9474(00)00372-9} {\bibfield  {journal} {\bibinfo  {journal}
  {Nucl. Phys. A}\ }\textbf {\bibinfo {volume} {679}},\ \bibinfo {pages} {462}
  (\bibinfo {year} {2001})}\BibitemShut {NoStop}%
\bibitem [{\citenamefont {Meister}\ \emph {et~al.}(2002)\citenamefont {Meister}
  \emph {et~al.}}]{Meister:2002uuu}%
  \BibitemOpen
  \bibfield  {author} {\bibinfo {author} {\bibfnamefont {M.}~\bibnamefont
  {Meister}} \emph {et~al.},\ }\href {\doibase 10.1016/S0375-9474(01)01305-7}
  {\bibfield  {journal} {\bibinfo  {journal} {Nucl. Phys. A}\ }\textbf
  {\bibinfo {volume} {700}},\ \bibinfo {pages} {3} (\bibinfo {year}
  {2002})}\BibitemShut {NoStop}%
\bibitem [{\citenamefont {Golovkov}\ \emph {et~al.}(2009)\citenamefont
  {Golovkov} \emph {et~al.}}]{Golovkov:2008kg}%
  \BibitemOpen
  \bibfield  {author} {\bibinfo {author} {\bibfnamefont {M.~S.}\ \bibnamefont
  {Golovkov}} \emph {et~al.},\ }\href {\doibase 10.1016/j.physletb.2008.12.052}
  {\bibfield  {journal} {\bibinfo  {journal} {Phys. Lett. B}\ }\textbf
  {\bibinfo {volume} {672}},\ \bibinfo {pages} {22} (\bibinfo {year}
  {2009})}\BibitemShut {NoStop}%
\bibitem [{\citenamefont {Grigorenko}\ \emph {et~al.}(2009)\citenamefont
  {Grigorenko} \emph {et~al.}}]{Grigorenko:2009}%
  \BibitemOpen
  \bibfield  {author} {\bibinfo {author} {\bibfnamefont {L.~V.}\ \bibnamefont
  {Grigorenko}} \emph {et~al.},\ }\href@noop {} {\bibfield  {journal} {\bibinfo
   {journal} {PEPAN}\ }\textbf {\bibinfo {volume} {6}},\ \bibinfo {pages} {118}
  (\bibinfo {year} {2009})}\BibitemShut {NoStop}%
\bibitem [{\citenamefont {Iwata}\ \emph {et~al.}(2000)\citenamefont {Iwata}
  \emph {et~al.}}]{Iwata:2000dd}%
  \BibitemOpen
  \bibfield  {author} {\bibinfo {author} {\bibfnamefont {Y.}~\bibnamefont
  {Iwata}} \emph {et~al.},\ }\href {\doibase 10.1103/PhysRevC.62.064311}
  {\bibfield  {journal} {\bibinfo  {journal} {Phys. Rev. C}\ }\textbf {\bibinfo
  {volume} {62}},\ \bibinfo {pages} {064311} (\bibinfo {year}
  {2000})}\BibitemShut {NoStop}%
\bibitem [{\citenamefont {Holl}\ \emph {et~al.}(2021)\citenamefont {Holl} \emph
  {et~al.}}]{Holl:2021bxg}%
  \BibitemOpen
  \bibfield  {author} {\bibinfo {author} {\bibfnamefont {M.}~\bibnamefont
  {Holl}} \emph {et~al.},\ }\href {\doibase 10.1016/j.physletb.2021.136710}
  {\bibfield  {journal} {\bibinfo  {journal} {Phys. Lett. B}\ }\textbf
  {\bibinfo {volume} {822}},\ \bibinfo {pages} {136710} (\bibinfo {year}
  {2021})}\BibitemShut {NoStop}%
\bibitem [{\citenamefont {Aumann}(2022)}]{Aumann:2022PC}%
  \BibitemOpen
  \bibfield  {author} {\bibinfo {author} {\bibfnamefont {T.}~\bibnamefont
  {Aumann}},\ }\href@noop {} {}\bibinfo {howpublished} {private communication}
  (\bibinfo {year} {2022})\BibitemShut {NoStop}%
\bibitem [{\citenamefont {Bacca}\ \emph {et~al.}(2009)\citenamefont {Bacca},
  \citenamefont {Schwenk}, \citenamefont {Hagen},\ and\ \citenamefont
  {Papenbrock}}]{Bacca:2009yk}%
  \BibitemOpen
  \bibfield  {author} {\bibinfo {author} {\bibfnamefont {S.}~\bibnamefont
  {Bacca}}, \bibinfo {author} {\bibfnamefont {A.}~\bibnamefont {Schwenk}},
  \bibinfo {author} {\bibfnamefont {G.}~\bibnamefont {Hagen}}, \ and\ \bibinfo
  {author} {\bibfnamefont {T.}~\bibnamefont {Papenbrock}},\ }\href {\doibase
  10.1140/epja/i2009-10815-5} {\bibfield  {journal} {\bibinfo  {journal} {Eur.
  Phys. J. A}\ }\textbf {\bibinfo {volume} {42}},\ \bibinfo {pages} {553}
  (\bibinfo {year} {2009})}\BibitemShut {NoStop}%
\bibitem [{\citenamefont {Caprio}\ \emph {et~al.}(2014)\citenamefont {Caprio},
  \citenamefont {Maris},\ and\ \citenamefont {Vary}}]{Caprio:2014iha}%
  \BibitemOpen
  \bibfield  {author} {\bibinfo {author} {\bibfnamefont {M.~A.}\ \bibnamefont
  {Caprio}}, \bibinfo {author} {\bibfnamefont {P.}~\bibnamefont {Maris}}, \
  and\ \bibinfo {author} {\bibfnamefont {J.~P.}\ \bibnamefont {Vary}},\ }\href
  {\doibase 10.1103/PhysRevC.90.034305} {\bibfield  {journal} {\bibinfo
  {journal} {Phys. Rev. C}\ }\textbf {\bibinfo {volume} {90}},\ \bibinfo
  {pages} {034305} (\bibinfo {year} {2014})}\BibitemShut {NoStop}%
\bibitem [{\citenamefont {Bonaiti}\ \emph {et~al.}(2021)\citenamefont
  {Bonaiti}, \citenamefont {Bacca},\ and\ \citenamefont
  {Hagen}}]{Bonaiti:2021kkp}%
  \BibitemOpen
  \bibfield  {author} {\bibinfo {author} {\bibfnamefont {F.}~\bibnamefont
  {Bonaiti}}, \bibinfo {author} {\bibfnamefont {S.}~\bibnamefont {Bacca}}, \
  and\ \bibinfo {author} {\bibfnamefont {G.}~\bibnamefont {Hagen}},\
  }\href@noop {} {\  (\bibinfo {year} {2021})},\ \Eprint
  {http://arxiv.org/abs/2112.08210} {arXiv:2112.08210 [nucl-th]} \BibitemShut
  {NoStop}%
\bibitem [{\citenamefont {Bacca}\ \emph {et~al.}()\citenamefont {Bacca},
  \citenamefont {Barnea}, \citenamefont {Hagen}, \citenamefont {Orlandini},\
  and\ \citenamefont {Papenbrock}}]{Bacca:2013dma}%
  \BibitemOpen
  \bibfield  {author} {\bibinfo {author} {\bibfnamefont {S.}~\bibnamefont
  {Bacca}}, \bibinfo {author} {\bibfnamefont {N.}~\bibnamefont {Barnea}},
  \bibinfo {author} {\bibfnamefont {G.}~\bibnamefont {Hagen}}, \bibinfo
  {author} {\bibfnamefont {G.}~\bibnamefont {Orlandini}}, \ and\ \bibinfo
  {author} {\bibfnamefont {T.}~\bibnamefont {Papenbrock}},\ }\href@noop {} {\
  }\BibitemShut {NoStop}%
\bibitem [{\citenamefont {Hohenberg}\ and\ \citenamefont
  {Kohn}(1964)}]{Hohenberg:1964zz}%
  \BibitemOpen
  \bibfield  {author} {\bibinfo {author} {\bibfnamefont {P.}~\bibnamefont
  {Hohenberg}}\ and\ \bibinfo {author} {\bibfnamefont {W.}~\bibnamefont
  {Kohn}},\ }\href {\doibase 10.1103/PhysRev.136.B864} {\bibfield  {journal}
  {\bibinfo  {journal} {Phys. Rev.}\ }\textbf {\bibinfo {volume} {136}},\
  \bibinfo {pages} {B864} (\bibinfo {year} {1964})}\BibitemShut {NoStop}%
\bibitem [{\citenamefont {Kohn}\ and\ \citenamefont {Sham}(1965)}]{Kohn:1965}%
  \BibitemOpen
  \bibfield  {author} {\bibinfo {author} {\bibfnamefont {W.}~\bibnamefont
  {Kohn}}\ and\ \bibinfo {author} {\bibfnamefont {L.~J.}\ \bibnamefont
  {Sham}},\ }\href {\doibase 10.1103/PhysRev.140.A1133} {\bibfield  {journal}
  {\bibinfo  {journal} {Phys. Rev.}\ }\textbf {\bibinfo {volume} {140}},\
  \bibinfo {pages} {A1133} (\bibinfo {year} {1965})}\BibitemShut {NoStop}%
\bibitem [{\citenamefont {Furnstahl}(2020)}]{Furnstahl:2019lue}%
  \BibitemOpen
  \bibfield  {author} {\bibinfo {author} {\bibfnamefont {R.~J.}\ \bibnamefont
  {Furnstahl}},\ }\href {\doibase 10.1140/epja/s10050-020-00095-y} {\bibfield
  {journal} {\bibinfo  {journal} {Eur. Phys. J. A}\ }\textbf {\bibinfo {volume}
  {56}},\ \bibinfo {pages} {85} (\bibinfo {year} {2020})}\BibitemShut {NoStop}%
\bibitem [{\citenamefont {Skyrme}(1956)}]{Skyrme:1956zz}%
  \BibitemOpen
  \bibfield  {author} {\bibinfo {author} {\bibfnamefont {T.~H.~R.}\
  \bibnamefont {Skyrme}},\ }\href {\doibase 10.1080/14786435608238186}
  {\bibfield  {journal} {\bibinfo  {journal} {Phil. Mag.}\ }\textbf {\bibinfo
  {volume} {1}},\ \bibinfo {pages} {1043} (\bibinfo {year} {1956})}\BibitemShut
  {NoStop}%
\bibitem [{\citenamefont {Skyrme}(1959)}]{Skyrme:1959zz}%
  \BibitemOpen
  \bibfield  {author} {\bibinfo {author} {\bibfnamefont {T.}~\bibnamefont
  {Skyrme}},\ }\href {\doibase 10.1016/0029-5582(58)90345-6} {\bibfield
  {journal} {\bibinfo  {journal} {Nucl. Phys.}\ }\textbf {\bibinfo {volume}
  {9}},\ \bibinfo {pages} {615} (\bibinfo {year} {1959})}\BibitemShut {NoStop}%
\bibitem [{\citenamefont {Serot}\ and\ \citenamefont
  {Walecka}(1986)}]{Serot:1984ey}%
  \BibitemOpen
  \bibfield  {author} {\bibinfo {author} {\bibfnamefont {B.~D.}\ \bibnamefont
  {Serot}}\ and\ \bibinfo {author} {\bibfnamefont {J.~D.}\ \bibnamefont
  {Walecka}},\ }\href@noop {} {\bibfield  {journal} {\bibinfo  {journal} {Adv.
  Nucl. Phys.}\ }\textbf {\bibinfo {volume} {16}},\ \bibinfo {pages} {1}
  (\bibinfo {year} {1986})}\BibitemShut {NoStop}%
\bibitem [{\citenamefont {Yang}\ and\ \citenamefont
  {Piekarewicz}(2020)}]{Yang:2019fvs}%
  \BibitemOpen
  \bibfield  {author} {\bibinfo {author} {\bibfnamefont {J.}~\bibnamefont
  {Yang}}\ and\ \bibinfo {author} {\bibfnamefont {J.}~\bibnamefont
  {Piekarewicz}},\ }\href {\doibase 10.1146/annurev-nucl-101918-023608}
  {\bibfield  {journal} {\bibinfo  {journal} {Ann. Rev. Nucl. Part. Sci.}\
  }\textbf {\bibinfo {volume} {70}},\ \bibinfo {pages} {21} (\bibinfo {year}
  {2020})}\BibitemShut {NoStop}%
\bibitem [{\citenamefont {Engel}(2007)}]{Engel:2006qu}%
  \BibitemOpen
  \bibfield  {author} {\bibinfo {author} {\bibfnamefont {J.}~\bibnamefont
  {Engel}},\ }\href {\doibase 10.1103/PhysRevC.75.014306} {\bibfield  {journal}
  {\bibinfo  {journal} {Phys. Rev. C}\ }\textbf {\bibinfo {volume} {75}},\
  \bibinfo {pages} {014306} (\bibinfo {year} {2007})}\BibitemShut {NoStop}%
\bibitem [{\citenamefont {Alex~Brown}(1998)}]{Brown:1998}%
  \BibitemOpen
  \bibfield  {author} {\bibinfo {author} {\bibfnamefont {B.}~\bibnamefont
  {Alex~Brown}},\ }\href {\doibase 10.1103/PhysRevC.58.220} {\bibfield
  {journal} {\bibinfo  {journal} {Phys. Rev. C}\ }\textbf {\bibinfo {volume}
  {58}},\ \bibinfo {pages} {220} (\bibinfo {year} {1998})}\BibitemShut
  {NoStop}%
\bibitem [{\citenamefont {Thouless}(1961)}]{Thouless:1961}%
  \BibitemOpen
  \bibfield  {author} {\bibinfo {author} {\bibfnamefont {D.}~\bibnamefont
  {Thouless}},\ }\href@noop {} {\bibfield  {journal} {\bibinfo  {journal}
  {Nuclear Physics}\ }\textbf {\bibinfo {volume} {22}},\ \bibinfo {pages} {78 }
  (\bibinfo {year} {1961})}\BibitemShut {NoStop}%
\bibitem [{\citenamefont {Mueller}\ and\ \citenamefont
  {Serot}(1996)}]{Mueller:1996pm}%
  \BibitemOpen
  \bibfield  {author} {\bibinfo {author} {\bibfnamefont {H.}~\bibnamefont
  {Mueller}}\ and\ \bibinfo {author} {\bibfnamefont {B.~D.}\ \bibnamefont
  {Serot}},\ }\href@noop {} {\bibfield  {journal} {\bibinfo  {journal} {Nucl.
  Phys.}\ }\textbf {\bibinfo {volume} {A606}},\ \bibinfo {pages} {508}
  (\bibinfo {year} {1996})}\BibitemShut {NoStop}%
\bibitem [{\citenamefont {Horowitz}\ and\ \citenamefont
  {Piekarewicz}(2001)}]{Horowitz:2000xj}%
  \BibitemOpen
  \bibfield  {author} {\bibinfo {author} {\bibfnamefont {C.~J.}\ \bibnamefont
  {Horowitz}}\ and\ \bibinfo {author} {\bibfnamefont {J.}~\bibnamefont
  {Piekarewicz}},\ }\href@noop {} {\bibfield  {journal} {\bibinfo  {journal}
  {Phys. Rev. Lett.}\ }\textbf {\bibinfo {volume} {86}},\ \bibinfo {pages}
  {5647} (\bibinfo {year} {2001})}\BibitemShut {NoStop}%
\bibitem [{\citenamefont {Chen}\ and\ \citenamefont
  {Piekarewicz}(2014)}]{Chen:2014sca}%
  \BibitemOpen
  \bibfield  {author} {\bibinfo {author} {\bibfnamefont {W.-C.}\ \bibnamefont
  {Chen}}\ and\ \bibinfo {author} {\bibfnamefont {J.}~\bibnamefont
  {Piekarewicz}},\ }\href@noop {} {\bibfield  {journal} {\bibinfo  {journal}
  {Phys. Rev.}\ }\textbf {\bibinfo {volume} {C90}},\ \bibinfo {pages} {044305}
  (\bibinfo {year} {2014})}\BibitemShut {NoStop}%
\bibitem [{\citenamefont {Dickhoff}\ and\ \citenamefont
  {Van~Neck}(2005)}]{Dickhoff:2005}%
  \BibitemOpen
  \bibfield  {author} {\bibinfo {author} {\bibfnamefont {W.~H.}\ \bibnamefont
  {Dickhoff}}\ and\ \bibinfo {author} {\bibfnamefont {D.}~\bibnamefont
  {Van~Neck}},\ }\enquote {\bibinfo {title} {Many-body theory exposed},}\ \
  (\bibinfo  {publisher} {World Scientific Publishing Co.},\ \bibinfo {year}
  {2005})\BibitemShut {NoStop}%
\bibitem [{\citenamefont {Piekarewicz}(2014)}]{Piekarewicz:2013bea}%
  \BibitemOpen
  \bibfield  {author} {\bibinfo {author} {\bibfnamefont {J.}~\bibnamefont
  {Piekarewicz}},\ }\href {\doibase 10.1140/epja/i2014-14025-x} {\bibfield
  {journal} {\bibinfo  {journal} {Eur. Phys. J. A}\ }\textbf {\bibinfo {volume}
  {50}},\ \bibinfo {pages} {25} (\bibinfo {year} {2014})}\BibitemShut {NoStop}%
\bibitem [{\citenamefont {Harakeh}\ and\ \citenamefont {van~der
  Woude}(2001)}]{Harakeh:2001}%
  \BibitemOpen
  \bibfield  {author} {\bibinfo {author} {\bibfnamefont {M.~N.}\ \bibnamefont
  {Harakeh}}\ and\ \bibinfo {author} {\bibfnamefont {A.}~\bibnamefont {van~der
  Woude}},\ }\enquote {\bibinfo {title} {Giant resonances-fundamental
  high-frequency modes of nuclear excitation},}\ \ (\bibinfo  {publisher}
  {Clarendon, Oxford},\ \bibinfo {year} {2001})\BibitemShut {NoStop}%
\bibitem [{\citenamefont {Piekarewicz}(2021)}]{Piekarewicz:2021jte}%
  \BibitemOpen
  \bibfield  {author} {\bibinfo {author} {\bibfnamefont {J.}~\bibnamefont
  {Piekarewicz}},\ }\href {\doibase 10.1103/PhysRevC.104.024329} {\bibfield
  {journal} {\bibinfo  {journal} {Phys. Rev. C}\ }\textbf {\bibinfo {volume}
  {104}},\ \bibinfo {pages} {024329} (\bibinfo {year} {2021})}\BibitemShut
  {NoStop}%
\bibitem [{\citenamefont {Roca-Maza}\ \emph {et~al.}(2013)\citenamefont
  {Roca-Maza}, \citenamefont {Centelles}, \citenamefont {Vi\~nas},
  \citenamefont {Brenna}, \citenamefont {Col\`o} \emph
  {et~al.}}]{Roca-Maza:2013mla}%
  \BibitemOpen
  \bibfield  {author} {\bibinfo {author} {\bibfnamefont {X.}~\bibnamefont
  {Roca-Maza}}, \bibinfo {author} {\bibfnamefont {M.}~\bibnamefont
  {Centelles}}, \bibinfo {author} {\bibfnamefont {X.}~\bibnamefont {Vi\~nas}},
  \bibinfo {author} {\bibfnamefont {M.}~\bibnamefont {Brenna}}, \bibinfo
  {author} {\bibfnamefont {G.}~\bibnamefont {Col\`o}},  \emph {et~al.},\ }\href
  {\doibase 10.1103/PhysRevC.88.024316} {\bibfield  {journal} {\bibinfo
  {journal} {Phys. Rev.}\ }\textbf {\bibinfo {volume} {C88}},\ \bibinfo {pages}
  {024316} (\bibinfo {year} {2013})}\BibitemShut {NoStop}%
\bibitem [{\citenamefont {Reinhard}\ and\ \citenamefont
  {Nazarewicz}(2010)}]{Reinhard:2010wz}%
  \BibitemOpen
  \bibfield  {author} {\bibinfo {author} {\bibfnamefont {P.-G.}\ \bibnamefont
  {Reinhard}}\ and\ \bibinfo {author} {\bibfnamefont {W.}~\bibnamefont
  {Nazarewicz}},\ }\href {\doibase 10.1103/PhysRevC.81.051303} {\bibfield
  {journal} {\bibinfo  {journal} {Phys. Rev.}\ }\textbf {\bibinfo {volume}
  {C81}},\ \bibinfo {pages} {051303(R)} (\bibinfo {year} {2010})}\BibitemShut
  {NoStop}%
\bibitem [{\citenamefont {Piekarewicz}\ \emph {et~al.}(2012)\citenamefont
  {Piekarewicz}, \citenamefont {Agrawal}, \citenamefont {Col\`o}, \citenamefont
  {Nazarewicz}, \citenamefont {Paar} \emph {et~al.}}]{Piekarewicz:2012pp}%
  \BibitemOpen
  \bibfield  {author} {\bibinfo {author} {\bibfnamefont {J.}~\bibnamefont
  {Piekarewicz}}, \bibinfo {author} {\bibfnamefont {B.}~\bibnamefont
  {Agrawal}}, \bibinfo {author} {\bibfnamefont {G.}~\bibnamefont {Col\`o}},
  \bibinfo {author} {\bibfnamefont {W.}~\bibnamefont {Nazarewicz}}, \bibinfo
  {author} {\bibfnamefont {N.}~\bibnamefont {Paar}},  \emph {et~al.},\ }\href
  {\doibase 10.1103/PhysRevC.85.041302} {\bibfield  {journal} {\bibinfo
  {journal} {Phys. Rev.}\ }\textbf {\bibinfo {volume} {C85}},\ \bibinfo {pages}
  {041302(R)} (\bibinfo {year} {2012})}\BibitemShut {NoStop}%
\bibitem [{\citenamefont {Chen}\ and\ \citenamefont
  {Piekarewicz}(2015)}]{Chen:2014mza}%
  \BibitemOpen
  \bibfield  {author} {\bibinfo {author} {\bibfnamefont {W.-C.}\ \bibnamefont
  {Chen}}\ and\ \bibinfo {author} {\bibfnamefont {J.}~\bibnamefont
  {Piekarewicz}},\ }\href@noop {} {\bibfield  {journal} {\bibinfo  {journal}
  {Phys. Lett.}\ }\textbf {\bibinfo {volume} {B748}},\ \bibinfo {pages} {284}
  (\bibinfo {year} {2015})}\BibitemShut {NoStop}%
\bibitem [{\citenamefont {Adhikari}\ \emph {et~al.}(2021)\citenamefont
  {Adhikari} \emph {et~al.}}]{Adhikari:2021phr}%
  \BibitemOpen
  \bibfield  {author} {\bibinfo {author} {\bibfnamefont {D.}~\bibnamefont
  {Adhikari}} \emph {et~al.} (\bibinfo {collaboration} {PREX}),\ }\href
  {\doibase 10.1103/PhysRevLett.126.172502} {\bibfield  {journal} {\bibinfo
  {journal} {Phys. Rev. Lett.}\ }\textbf {\bibinfo {volume} {126}},\ \bibinfo
  {pages} {172502} (\bibinfo {year} {2021})}\BibitemShut {NoStop}%
\bibitem [{\citenamefont {Brodeur}\ \emph {et~al.}(2012)\citenamefont {Brodeur}
  \emph {et~al.}}]{Brodeur:2011sam}%
  \BibitemOpen
  \bibfield  {author} {\bibinfo {author} {\bibfnamefont {M.}~\bibnamefont
  {Brodeur}} \emph {et~al.},\ }\href {\doibase 10.1103/PhysRevLett.108.052504}
  {\bibfield  {journal} {\bibinfo  {journal} {Phys. Rev. Lett.}\ }\textbf
  {\bibinfo {volume} {108}},\ \bibinfo {pages} {052504} (\bibinfo {year}
  {2012})}\BibitemShut {NoStop}%
\bibitem [{\citenamefont {Huang}\ \emph {et~al.}(2021)\citenamefont {Huang},
  \citenamefont {Wang}, \citenamefont {Kondev}, \citenamefont {Audi},\ and\
  \citenamefont {Naimi}}]{Huang:2021nwk}%
  \BibitemOpen
  \bibfield  {author} {\bibinfo {author} {\bibfnamefont {W.~J.}\ \bibnamefont
  {Huang}}, \bibinfo {author} {\bibfnamefont {M.}~\bibnamefont {Wang}},
  \bibinfo {author} {\bibfnamefont {F.~G.}\ \bibnamefont {Kondev}}, \bibinfo
  {author} {\bibfnamefont {G.}~\bibnamefont {Audi}}, \ and\ \bibinfo {author}
  {\bibfnamefont {S.}~\bibnamefont {Naimi}},\ }\href {\doibase
  10.1088/1674-1137/abddb0} {\bibfield  {journal} {\bibinfo  {journal} {Chin.
  Phys. C}\ }\textbf {\bibinfo {volume} {45}},\ \bibinfo {pages} {030002}
  (\bibinfo {year} {2021})}\BibitemShut {NoStop}%
\bibitem [{\citenamefont {Wang}\ \emph {et~al.}(2021)\citenamefont {Wang},
  \citenamefont {Huang}, \citenamefont {Kondev}, \citenamefont {Audi},\ and\
  \citenamefont {Naimi}}]{Wang:2021xhn}%
  \BibitemOpen
  \bibfield  {author} {\bibinfo {author} {\bibfnamefont {M.}~\bibnamefont
  {Wang}}, \bibinfo {author} {\bibfnamefont {W.~J.}\ \bibnamefont {Huang}},
  \bibinfo {author} {\bibfnamefont {F.~G.}\ \bibnamefont {Kondev}}, \bibinfo
  {author} {\bibfnamefont {G.}~\bibnamefont {Audi}}, \ and\ \bibinfo {author}
  {\bibfnamefont {S.}~\bibnamefont {Naimi}},\ }\href {\doibase
  10.1088/1674-1137/abddaf} {\bibfield  {journal} {\bibinfo  {journal} {Chin.
  Phys. C}\ }\textbf {\bibinfo {volume} {45}},\ \bibinfo {pages} {030003}
  (\bibinfo {year} {2021})}\BibitemShut {NoStop}%
\bibitem [{\citenamefont {Mueller}\ \emph {et~al.}(2007)\citenamefont {Mueller}
  \emph {et~al.}}]{Mueller:2007dhq}%
  \BibitemOpen
  \bibfield  {author} {\bibinfo {author} {\bibfnamefont {P.}~\bibnamefont
  {Mueller}} \emph {et~al.},\ }\href {\doibase 10.1103/PhysRevLett.99.252501}
  {\bibfield  {journal} {\bibinfo  {journal} {Phys. Rev. Lett.}\ }\textbf
  {\bibinfo {volume} {99}},\ \bibinfo {pages} {252501} (\bibinfo {year}
  {2007})}\BibitemShut {NoStop}%
\bibitem [{\citenamefont {Liu}\ \emph {et~al.}(2021)\citenamefont {Liu},
  \citenamefont {Egelhof}, \citenamefont {Kiselev},\ and\ \citenamefont
  {Mutterer}}]{Liu:2021cbn}%
  \BibitemOpen
  \bibfield  {author} {\bibinfo {author} {\bibfnamefont {X.}~\bibnamefont
  {Liu}}, \bibinfo {author} {\bibfnamefont {P.}~\bibnamefont {Egelhof}},
  \bibinfo {author} {\bibfnamefont {O.}~\bibnamefont {Kiselev}}, \ and\
  \bibinfo {author} {\bibfnamefont {M.}~\bibnamefont {Mutterer}},\ }\href
  {\doibase 10.1103/PhysRevC.104.034315} {\bibfield  {journal} {\bibinfo
  {journal} {Phys. Rev. C}\ }\textbf {\bibinfo {volume} {104}},\ \bibinfo
  {pages} {034315} (\bibinfo {year} {2021})}\BibitemShut {NoStop}%
\bibitem [{\citenamefont {Krauth}\ \emph {et~al.}(2021)\citenamefont {Krauth}
  \emph {et~al.}}]{Krauth:2021foz}%
  \BibitemOpen
  \bibfield  {author} {\bibinfo {author} {\bibfnamefont {J.~J.}\ \bibnamefont
  {Krauth}} \emph {et~al.},\ }\href {\doibase 10.1038/s41586-021-03183-1}
  {\bibfield  {journal} {\bibinfo  {journal} {Nature}\ }\textbf {\bibinfo
  {volume} {589}},\ \bibinfo {pages} {527} (\bibinfo {year}
  {2021})}\BibitemShut {NoStop}%
\bibitem [{\citenamefont {Horowitz}\ and\ \citenamefont
  {Piekarewicz}(2012)}]{Horowitz:2012we}%
  \BibitemOpen
  \bibfield  {author} {\bibinfo {author} {\bibfnamefont {C.~J.}\ \bibnamefont
  {Horowitz}}\ and\ \bibinfo {author} {\bibfnamefont {J.}~\bibnamefont
  {Piekarewicz}},\ }\href {\doibase 10.1103/PhysRevC.86.045503} {\bibfield
  {journal} {\bibinfo  {journal} {Phys.Rev.}\ }\textbf {\bibinfo {volume}
  {C86}},\ \bibinfo {pages} {045503} (\bibinfo {year} {2012})}\BibitemShut
  {NoStop}%
\bibitem [{\citenamefont {Thiel}\ \emph {et~al.}(2019)\citenamefont {Thiel},
  \citenamefont {Sfienti}, \citenamefont {Piekarewicz}, \citenamefont
  {Horowitz},\ and\ \citenamefont {Vanderhaeghen}}]{Thiel:2019tkm}%
  \BibitemOpen
  \bibfield  {author} {\bibinfo {author} {\bibfnamefont {M.}~\bibnamefont
  {Thiel}}, \bibinfo {author} {\bibfnamefont {C.}~\bibnamefont {Sfienti}},
  \bibinfo {author} {\bibfnamefont {J.}~\bibnamefont {Piekarewicz}}, \bibinfo
  {author} {\bibfnamefont {C.~J.}\ \bibnamefont {Horowitz}}, \ and\ \bibinfo
  {author} {\bibfnamefont {M.}~\bibnamefont {Vanderhaeghen}},\ }\href {\doibase
  10.1088/1361-6471/ab2c6d} {\bibfield  {journal} {\bibinfo  {journal} {J.
  Phys.}\ }\textbf {\bibinfo {volume} {G46}},\ \bibinfo {pages} {093003}
  (\bibinfo {year} {2019})}\BibitemShut {NoStop}%
\bibitem [{\citenamefont {Mihaila}\ and\ \citenamefont
  {Heisenberg}(1999)}]{Mihaila:1998qr}%
  \BibitemOpen
  \bibfield  {author} {\bibinfo {author} {\bibfnamefont {B.}~\bibnamefont
  {Mihaila}}\ and\ \bibinfo {author} {\bibfnamefont {J.~H.}\ \bibnamefont
  {Heisenberg}},\ }\href {\doibase 10.1103/PhysRevC.60.054303} {\bibfield
  {journal} {\bibinfo  {journal} {Phys. Rev. C}\ }\textbf {\bibinfo {volume}
  {60}},\ \bibinfo {pages} {054303} (\bibinfo {year} {1999})}\BibitemShut
  {NoStop}%
\bibitem [{\citenamefont {Piekarewicz}(2001)}]{Piekarewicz:2001nm}%
  \BibitemOpen
  \bibfield  {author} {\bibinfo {author} {\bibfnamefont {J.}~\bibnamefont
  {Piekarewicz}},\ }\href@noop {} {\bibfield  {journal} {\bibinfo  {journal}
  {Phys. Rev.}\ }\textbf {\bibinfo {volume} {C64}} (\bibinfo {year}
  {2001})}\BibitemShut {NoStop}%
\end{thebibliography}%

\end{document}